\documentclass[usenatbib,usegraphicx,twocolumn]{mn2e}

\newcommand{\be}{\begin{equation}}
\newcommand{\e}{\end{equation}}
\newcommand{\bear}{\begin{eqnarray}}
\newcommand{\ear}{\end{eqnarray}}

\def\I{I_{\nu}}

\def\mnras{MNRAS}
\def\apj{Ap.J}

\def\apjs{Ap.JS}

\def\aap{A \& A}
\def\u{{\bf U}} 
\def\th{\vec{\theta}}
\def\V{\mathcal{V}}

\def\del{\partial}
\usepackage{graphics}
\begin{document}
\date {}
\title[GMRT observations : Improved 21-cm foreground removal]
{Improved foreground removal in  GMRT $610 \,{\rm MHz}$ observations 
towards redshifted 21-cm tomography}

\author[A. Ghosh, S. Bharadwaj, S. S. Ali and J. N. Chengalur]{Abhik
  Ghosh$^{1}$\thanks{E-mail: abhik@phy.iitkgp.ernet.in}, Somnath
  Bharadwaj$^{1}$\thanks{Email:somnath@phy.iitkgp.ernet.in},
  Sk. Saiyad Ali$^{2}$\thanks{Email:saiyad@phys.jdvu.ac.in} and \and
  Jayaram N. Chengalur$^{3}$\thanks{Email:chengalu@ncra.tifr.res.in}
  \\$^{1}$ Department of Physics and Meteorology \& Centre for
  Theoretical Studies , IIT Kharagpur, 721 302 , India \\$^{2}$
  Department of Physics,Jadavpur University, Kolkata 700032, India.
  \\$^3$ National Centre for Radio Astrophysics, TIFR, Post Bag 3,
  Ganeshkhind, Pune 411 007, India}

\maketitle
\begin{abstract}

Foreground removal is a challenge for 21-cm tomography of the high
redshift Universe. We use archival GMRT data (obtained for completely
different astronomical goals) to estimate the foregrounds at a
redshift $\sim 1$. The statistic we use is the cross power spectrum
between two frequencies separated by $\Delta\nu$ at the angular
multipole $\ell$, or equivalently the multi-frequency angular power
spectrum $C_{\ell}(\Delta \nu)$. An earlier measurement of
$C_{\ell}(\Delta \nu)$ using this data had revealed the presence of
oscillatory patterns along $\Delta \nu$, which turned out to be a
severe impediment for foreground removal \citep{ghosh}.  Using the
same data, in this paper we show that it is possible to considerably
reduce these oscillations by suppressing the sidelobe response of the
primary antenna elements.  The suppression works best at the angular
multipoles $\ell$ for which there is a dense sampling of the u-v
plane. For three angular multipoles $\ell= 1405, 1602 \ {\rm
  and}\ 1876$, this sidelobe suppression along with a low order
polynomial fitting completely results in residuals of $(\le 0.02 \,
{\rm mK}^2)$, consistent with the noise at the $3\sigma$ level. Since
the polynomial fitting is done after estimation of the power spectrum
it can be ensured that the estimation of the HI signal is not
biased. The corresponding $99\%$ upper limit on the HI signal is
$\bar{x}_{\rm HI} b \le 2.9$, where $\bar{x}_{\rm HI}$ is the mean
neutral fraction and $b$ is the bias.

\end{abstract}

\section{Introduction}

Observations of redshifted 21 cm radiation from neutral hydrogen (HI)
hold the potential of tracing the large scale structure of the
Universe over a large redshift range $ (20 \ge z \ge 0) $. This signal
$(\sim \, {\rm mK})$, is however, buried in the emission from other
astrophysical sources which are collectively referred to as
foregrounds. These foregrounds are dominated by the extragalactic
radio sources, the diffuse synchrotron radiation from our own Galaxy
(GDSE) and a smaller contribution comes from galactic free-free
emission \citep{shaver}. The extragalactic point sources have a
typical spectral index of $\beta \sim -0.8$ \citep{Subrahmanyan}, with
evidence of flattening at lower frequencies \citep{cohen}. The
analysis of radio surveys at 408 MHz, 1.42 GHz, and 2.326 GHz
\citep*{haslam,R82,RR88,JBN98} show that the GDSE has a steep spectral
index to be ${\alpha} \approx -2.8$. This is also in broad agreement
with the results presented in \cite{platania1}.  \citet{dmat} have
used the 6C survey \citep{hales}, and the 3CR survey and the 3 CRR
catalogue \citep{laing} to estimate the resolved extragalactic radio
sources (point sources) contribution at $150 \, {\rm MHz}$. Recently
\citet{bernardi} have characterized the power spectrum of the total
diffuse radiation at $150 \, {\rm MHz}$ at the angular scales of our
interest. Separating the redshifted HI signal from the foregrounds,
which are several order of magnitude larger
(e.g. \citealt{shaver,dmat}), is currently the biggest challenge for
21-cm tomography of the high redshift Universe.  The problem, in
principle, can be solved using the fact that the redshifted 21-cm
signals at two different frequencies $\nu$ and $\nu \, + \, \Delta
\nu$ are expected to be uncorrelated at separations $\Delta \nu \ge
0.5 \, {\rm MHz}$ at the angular scales of our interest whereas the
foregrounds, which arise from continuum sources, are expected to
remain correlated over considerably large frequency separations
\citep{BSS,dmat,Oh,dmat04,zald,Mor04,BA5,Santos,wang}.  We note that
the prospects of detecting the redshifted 21-cm HI signal are
considerably better at higher frequencies (e.g., $610 \, {\rm MHz}$)
which probe the post-reionization era $(z < 6)$ in comparison to the
lower frequencies (e.g., $150 \, {\rm MHz}$ ) which probe the
reionization era.  Further, the problem of man made radio frequency
interference (hereafter RFI) is considerably less severe at higher
frequencies.

In this paper we report a substantial improvement in foreground
removal in the context of our earlier work (\citealt{ghosh}, hereafter
Paper I) which presents Giant Metrewave Radio Telescope
(GMRT\footnote{http://www.gmrt.ncra.tifr.res.in} \citealt{swarup})
$610 \, {\rm MHz}$ observations towards detecting the
post-reionization 21-cm signal from $z=1.32$. In Paper I we have
determined, possibly for the first time, the statistical properties of
the background radiation over the angular scales $20^{''}$ to $10^{'}$
and a frequency band of $7.5 \, {\rm MHz}$ centered at $616.25 \, {\rm
  MHz}$.  The analysis was carried out using the multi-frequency
angular power spectrum (MAPS) $C_{\ell}(\Delta \nu)$ \citep{kanan}
which jointly characterizes the angular $\ell$ and frequency $\Delta
\nu$ dependence of the fluctuations in the background radiation.  The
measured $C_{\ell}(\Delta \nu)$, which ranges from $7 \ {\rm mK}^2$ to
$18 \ {\rm mK^2}$, is dominated by foregrounds, the expected HI signal
being several orders of magnitude smaller ($C^{\rm HI}_{\ell}(\Delta
\nu) \sim 10^{-6} - 10^{-7} \ {\rm mK}^2$).  The measured signal, for
a fixed $\ell$, is expected to vary smoothly with $\Delta \nu$ and
remain nearly constant over the observational bandwidth. We find
instead that in addition to a component that exhibits a smooth $\Delta
\nu$ dependence, the measured $C_{\ell}(\Delta \nu)$ also has a
component that oscillates as a function of $\Delta \nu$.  The
amplitude of the oscillating component is around $1-4 \%$ of the
smooth component, and the amplitude and period of oscillation both
decreases with increasing $\ell$.  We note that similar oscillations,
with considerably larger amplitudes, have been reported in GMRT
observations at $153 \, {\rm MHz}$ \citep{ali} which is relevant for
the signal from the reionization era. The origin of the oscillatory
signal was unclear.
 
The oscillatory patterns pose a serious obstacle for foreground
removal (Paper I).  It is thus important to identify the cause (or
causes) and implement techniques to mitigate the oscillatory patterns.
The fact that the primary beam (hereafter PB) pattern changes with
frequency across the observational bandwidth, not included in our
previous analysis (Paper I), could be a possible cause.  In
particular, the angular position of the nulls and the side-lobes
changes with frequency, and a bright continuum source located near the
null or located in the sidelobes will be seen as oscillations along
the frequency axis in the measured visibilities.  It is thus quite
plausible that bright sources located near the null or the sidelobes
of the PB produce the oscillatory pattern in the measured
$C_{\ell}(\Delta \nu)$ which is estimated from correlations amongst
the visibilities.  One can, in principle, design antennas with a
frequency-dependent collecting area to produce a nearly constant PB
pattern.  However, to our knowledge, none of the upcoming arrays have
this feature and we expect this issue to be relevant not only for the
GMRT but for all the arrays planned in the near future.

The problem can be mitigated by tapering the array's sky response with
a frequency independent window function $W(\vec{\theta})$ that falls
off before the first null of the PB pattern and thereby suppresses the
sidelobe response.  It is simplest to implement this by multiplying
the sky image $I(\vec{\theta})$ with $W(\vec{\theta})$ and using a
Fourier transform of this to recalculate the visibilities.  This,
however, will introduce correlations between the noise in the
recalculated visibilities which is a nuisance for estimating
$C_{\ell}(\Delta \nu)$.  The other option is to work entirely with the
visibilities in the $u-v$ plane, avoiding the need for an image. In
this approach the sky response is tapered by convolving the
visibilities with $\tilde{W}(\vec{U})$ - the Fourier transform of
$W(\vec{\theta})$. As we show later in this paper, it is possible to
implement the convolution without introducing a noise bias in the
estimated $C_{\ell}(\Delta \nu)$. We note that convolving the
visibilities with a "gridding convolution function" has long been a
standard practice while making images from interferometric data
(Section 3., \citealt{Sramek}). This convolution is done to avoid
aliasing which would otherwise occur when one uses an FFT to make the
image. As discussed in detail by \citet{Sramek} the convolution
function is generally chosen to provide an optimum balance between
alias rejection and ease of computation. The focus here is somewhat
different, viz. to use the convolution to strongly attenuate the
frequency dependent response to the sidelobes of the primary antenna
pattern.

RFI sources, which are mostly located on the ground, are picked up
through the sidelobes.  Suppressing the sidelobe response is also
expected to mitigate the RFI contribution.

Paper I contains a detailed description of the data, we mention a few
salient features here. The data is taken from an archival 30 hours observation
centered on $\alpha_{2000}=12^h36^m49^s$, 
$\delta_{2000}= 62^{\circ}17^{'}57^{''}\, $ which is situated near
Hubble Deep Field North (HDF-N) .  For the present work we have
analyzed the frequency range $612.5 \,{\rm MHz}$ to $620.0 \,{\rm
  MHz}$ with channels $125 \,{\rm kHz}$ wide.  Visibilities were
recorded for two orthogonal circular polarizations with $16 {\rm s}$
integration time. Calibration was carried out using standard AIPS
tasks, and the visibilities from the two polarizations were combined
($\V=[\V_{RR}+\V_{LL}]/2$) for the rest of the analysis.

\section{Sidelobe Suppression}
\label{sec2}
The observed visibilities $\V=\tilde{a} \otimes\tilde{I} $  record 
 the  Fourier transform of  the sky brightness $\tilde{I}$ 
convolved with   the antenna  aperture $\tilde{a}$.  The frequency
dependence and the sidelobes come in through $\tilde{a}(\u,\nu)$
whose Fourier transform gives  the  PB pattern $A(\th,\nu)$.  Close
to the phase centre, the PB is   reasonably well modeled by a Gaussian 
$A(\th,\nu)=e^{-\theta^2/\theta_0^2}$  where the parameter $\theta_0$
is related to the FWHM of the PB as $\theta_0 \approx 0.6 \times
\theta_{\rm   FWHM}$,   and $\theta_0  =25^{'}.8$  ($\theta_{\rm
  FWHM}=43^{'}$) at 610 MHz  for the GMRT with $\theta_0 \propto
\nu^{-1}$.   Note that the Gaussian model for $A(\th,\nu)$  breaks
down away from the phase center where we have the sidelobes and
nulls.

We mitigate the effect of the frequency dependent sidelobe pattern by
convolving the observed visibilities with a suitably chosen function
$\tilde{W}(\u)$.  The sky response of the convolved visibilities
$\V_c=\tilde{W} \otimes \V$ is modulated by the window function
$W(\th)$ which is the Fourier transform of $\tilde{W}(\u)$.  We have
used a window function $W(\th)=e^{-\theta^2/\theta^2_w}$ with
$\theta_w < \theta_0$ to taper the sky response so that it falls off
well before the sidelobes.  Note that $\theta_w$ and $W(\th)$ are both
frequency independent. We parametrize $\theta_w$ as $\theta_w=\rm
{f}\theta_0$ with $\rm{f} \leq 1$ where $\theta_0$ here refers to the
value at the fixed frequency $610 \, {\rm MHz}$.

We can evaluate the convolved visibilities $\V_c$ on a grid in $u-v$
space ($\V_c(\u_i,\nu)=\sum_{a} \tilde W(\u_i - \u_{a}) \V
(\u_{a},\nu)$) and use these to determine the two-visibility
correlation defined as $ V_{2}(\u_i,\Delta\nu)=\V_c(\u_i,\nu)
\V_c^{*}(\u_i,\nu+ \Delta \nu)$.  Here $\u \equiv (u,v)$ refers to a
two-dimensional baseline, $\u_i$ refers to points on the grid in $u-v$
space and $\u_a$ refers to the different baselines in the
observational data.  This way of estimating $V_{2}(\u_i,\Delta\nu)$,
however, introduces a positive noise bias (e.g. \citet{Begum}) which
is not desirable.  We use, instead, the estimator
$V_{2}(\u_i,\Delta\nu)=K^{-1} \times \sum_{a\neq b} \left[ \tilde
  W(\u_i - \u_{a}) \tilde W^{*}(\u_i - \u_{b})\V
  (\u_{a},\nu)\V^{*}(\u_{b},\nu + \Delta\nu) \right]$, where
$K=\sum_{a\neq b}\tilde W(\u_i - \u_{a}) \tilde W^{*}(\u_i - \u_{b})$
is a normalization constant.  The noise bias is avoided by dropping
the self-correlations ({\it i.e..} the terms with $a=b$).  We have
used a grid of spacing $\Delta U_g=\sqrt{\ln 2}/(\pi \theta_w)=0.265
\theta_w^{-1}$ which corresponds to half of the FWHM of
$\tilde{W}(\u)$, and we have estimated $V_{2}(\u_i,\Delta\nu) $ at
every grid point using all the baselines within a disk of radius $2 \,
\Delta U_g$ centered on that grid point.  We finally determine
$C_{\ell}(\Delta\nu)$ using \citep{ali} \be V_{2}(U,\Delta
\nu)=\frac{\pi\theta'^{2}}{2} \left(\frac{\del \I}{\del T}\right)^2
C_{\ell}(\Delta \nu) \,Q(\Delta \nu) \,.
\label{eq:cell}
\e where $\ell= 2 \pi U$, $\theta'^{-2} = \theta_0 ^{-2} +
\theta_w^{-2}$ and $Q(\Delta \nu)$ is a slowly varying function of
$\Delta\nu$ which accounts for the fact that we have treated
$\theta'^{2} $ and $(\del \I/\del T) $ as constants in our analysis.
We have used $Q(\Delta \nu)=1$ which introduces an extra $\Delta \nu$
dependence in the estimated $C_{\ell}(\Delta \nu)$ .  This, we assume,
will be a small effect and can be accounted for during foreground
removal.  The data has been binned assuming that the statistical
properties of the signal are isotropic in $\u$.

\subsection{Simulation}

In order to demonstrate the efficacy of our technique of sidelobe
suppression using a simulated data set, the GMRT antenna was modeled
as a circular aperture of $D=45{\rm \, m}$ diameter with a circular
disk of diameter $D_1=1{\rm \, m}$ at the center blocked due to the
feed. This gives a normalized antenna beam pattern
\begin{equation}
A_{\nu}(\theta)=\frac{4}{(D^2-D_1^2)^2}\left[D^2 \frac{J_{1}(\pi \theta
    D/\lambda)}{(\pi \theta D/\lambda)} - D_1^2 \frac{J_{1}(\pi \theta
    D_1/\lambda)}{(\pi \theta 
    D_1/\lambda)}\right]^2
\label{beam}
\end{equation}
where $J_1$ is the Bessel function of the first kind of order one. 
Several point sources were randomly placed in the region near 
fifth null of this beam pattern. The expected number of point sources and
their flux distribution can be predicted from the differential source
counts \citep{garn}.  Given that the present simulation has the limited aim of
demonstrating the efficiency of our  sidelobe suppression technique, 
we have instead used a cartoon model with five hundred point sources 
each having the same flux density $150 \, {\rm mJy}$. The value of the 
flux density was chosen so that the simulated $C_{\ell}(\Delta \nu)$ has 
a value around $\sim 10 \, {\rm mK}^2$, comparable to our measured 
$C_{\ell}(\Delta \nu)$ (Figure 7, Paper I).

Next, we have randomly generated baselines, and calculated the
corresponding visibilities for the five-hundred point sources. It may 
be noted that the simulated baselines have a density of $0.3/{\rm m^2}$ 
which is roughly consistent with the GMRT central square.  The antenna beam
pattern (eq. \ref{beam}) was used in calculating the simulated
visibilities. We have used the simulated visibilities to estimate 
$C_{\ell}(\Delta \nu)$ . Then, we have fitted a third order polynomial 
in $\Delta \nu$ to subtract out the component of the sky signal that
varies slowly with frequency. The residual $C_{\ell}(\Delta \nu)$ has an 
oscillatory pattern with amplitude $(\sim 10^{-2} \, {\rm   mK}^2)$.
The procedure was repeated after introducing a convolution with $f=0.8$. 
This tapers the sky response to $\sim 0.8 \% $ of its peak value at the 
first null of the beam and considerably reduces the response to the 
sources beyond the first null.  We find that the residuals, after polynomial
subtraction,  are in the range  $10^{-5} \, {\rm mK}^2$ to  $10^{-4}
\, {\rm mK}^2$, i.e. the residuals are suppressed by a factor
of around $10^{2}$ to $10^{3}$. Further, after convolution the residuals 
appear to be noise like  and do not show any noticeable oscillatory feature.
Note that although we have shown results of our simulation for only a
single $\ell$ value, the results are very similar at other $\ell$
values.

\begin{figure*}
\includegraphics[width=70mm,angle=-90]{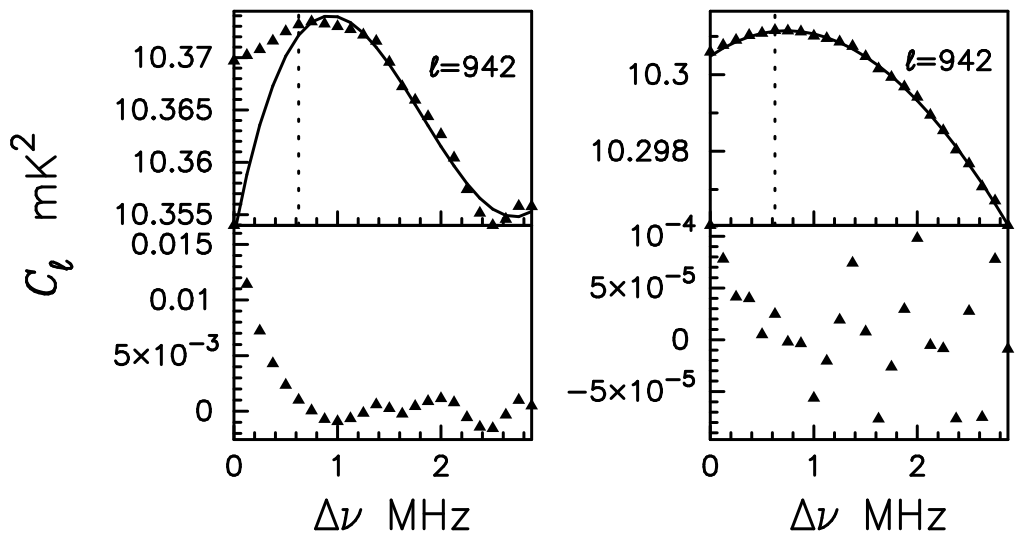}
\caption{The data points in the top panels show $C_{\ell}(\Delta\nu)$
  estimated without (left) and with (right) sidelobe suppression.  The
  solid lines show the respective best fit polynomials used for
  foreground subtraction.  The residuals, after foreground
  subtraction, are shown in the corresponding bottom panels.}
\label{clsimu}
\end{figure*}

\section{Results and Conclusions}
\label{sec3}
\begin{figure*}
\includegraphics[width=70mm,angle=-90]{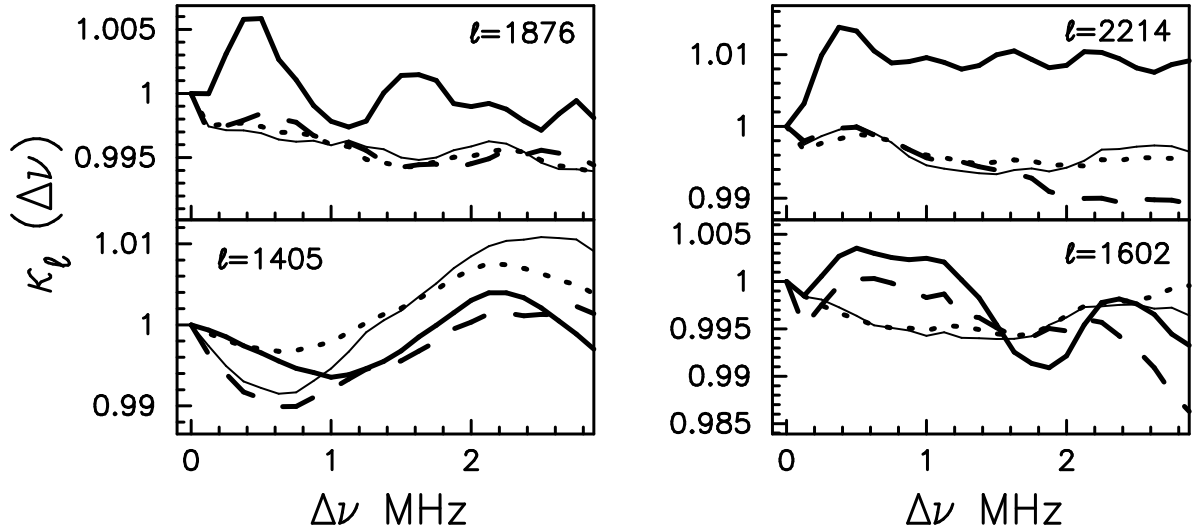}
\caption{The measured $\kappa_{\ell}(\Delta \nu)$ as a function of
  $\Delta \nu$ for the   $\ell$ value as shown in each
  panel.  The results from our earlier analysis (Paper I) and for a 
  tapering  ${\rm f}=0.4, 0.65, 0.8$ are shown by the  thick solid,
  dashed, thin solid  and dotted curves respectively. }
\label{kappa}
\end{figure*}

We first investigate whether sidelobe suppression at all helps to
reduce the oscillations that were reported in Paper I.  This is
quantified using the dimensionless decorrelation function
$\kappa_{\ell}(\Delta \nu)=C_{\ell}(\Delta \nu)/C_{\ell}(0)$.  We have
considered ${\rm f}=0.4, 0.65$ and $0.8$ which respectively correspond
to a tapered sky response with FWHM $17^{'}.2 , 28^{'}.0$ and
$34^{'}.4$ as compared to the GMRT PB which has a FWHM of $43^{'}$ at
$610 \, {\rm MHz}$.  The results are shown in Figure \ref{kappa} for
the four smallest $\ell$ values for which we have measured the binned
$C_{\ell}(\Delta \nu)$ .  The oscillatory patterns are distinctly
visible in the cases where the tapering has not been applied.  We see
that in most cases the oscillations are considerably reduced and are
nearly absent when tapering is applied.  We do not, however, notice
any particular qualitative trend with varying ${\rm f}$. The
oscillatory pattern with a large period in $\Delta \nu$ seen at
$\ell=1405$, the smallest $\ell$ value, persists even after the
tapering is applied. This, however, does not pose a problem for
foreground removal as the $C_{\ell}(\Delta \nu)$, after tapering, is
well fitted by a low order polynomial and can be successfully removed.

The tapering, which has been implemented through a convolution, is
expected to be most effective in a situation  where the $u-v$ space is
densely sampled by the baseline distribution.  Our results are 
limited by the patchy $u-v$ coverage of the present  observational
data.   This also possibly explains why the oscillations persist after
tapering at the smallest $\ell$ value.  Tapering the the field of view 
has a drawback in that this increases
the cosmic variance which scales as ${\rm f}^{-1}$.  It is possible to
compensate for this by repeating the entire analysis after adding
phases to the visibilities so as to shift the center of the field of
view.  We have not attempted this here, and our entire analysis is
restricted to a single field of view.  The increase in the cosmic
variance can also be used as a guiding principle in choosing the value
of ${\rm f}$. It is most advantageous to use the smallest value of
${\rm f}$ where the oscillations are adequately suppressed.  Of the
three values of ${\rm f}$ that we have considered here, we find that
foreground subtraction is most effective for ${\rm f}=0.65$, and we
use this for the entire subsequent analysis.

\begin{figure*}
\includegraphics[width=70mm,angle=-90]{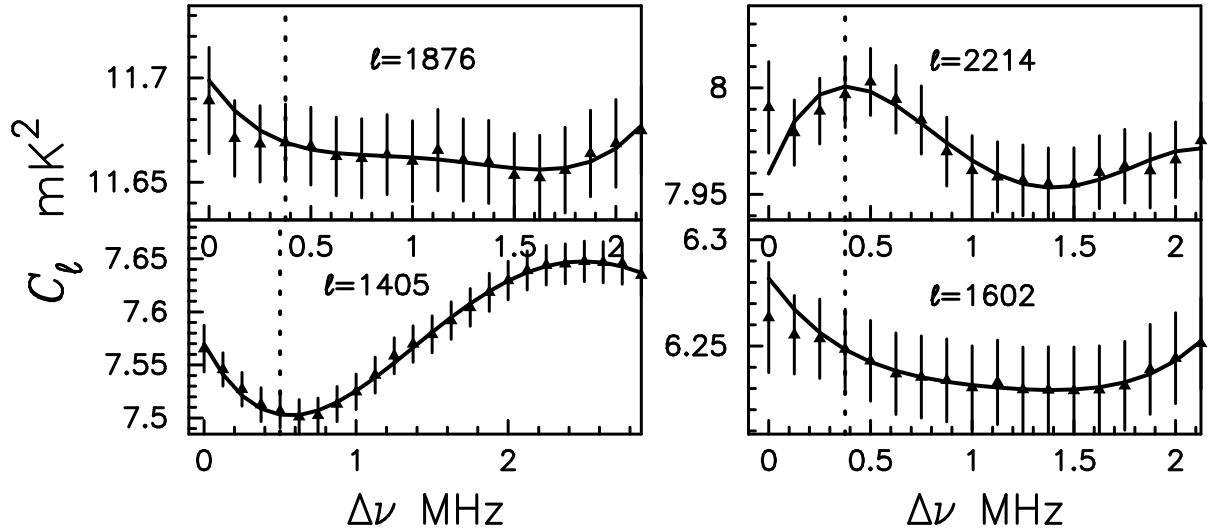}
\caption{The data points show the measured $C_{\ell}(\Delta \nu)$ as a
  function of $\Delta \nu$ for the $\ell$ value shown in each
  panel. The error-bars indicate the $3 \sigma$ system noise. The
  solid curve shows the $4^{th}$ order polynomial fits and the dotted
  vertical line shows $\Delta \nu_{0.1}$.}
\label{clnu}
\end{figure*}

The measured $C_{\ell}(\Delta \nu)$ shown in Figure \ref{clnu} is
foreground dominated and it is seen to vary smoothly with increasing
$\Delta \nu$.  On the contrary, the predicted contribution from the HI
signal $C^{\rm HI}_{\ell}(\Delta \nu)$ decreases very rapidly with
increasing $\Delta \nu$ \citep{BSS}.  We have used a polynomial
fitting technique (Paper I) to identify and subtract out any smoothly
varying component from the measured $C_{\ell}(\Delta \nu)$ and thereby
remove the foreground contribution.  The possibility that along with
the foregrounds the fitting procedure may also remove a part of the
signal is a major concern.  For each $\ell$ value we have estimated
$\Delta \nu_{0.1}$ which corresponds to the frequency separation where
$C^{\rm HI}_{\ell}(\Delta \nu)$ first falls to less than $10 \, \%$ of
the peak value $C^{\rm HI}_{\ell}(0)$.  The bulk of the HI signal is
localized within $\Delta \nu < \Delta \nu_{0.1}$ which is excluded
when estimating the slowly varying foreground contribution.  We have
used the range $ \Delta \nu_{0.1} \, \le \, \Delta \nu \le \, 6 \times
\Delta \nu_{0.1}$ to estimate the coefficients of the best fit
$4^{th}$ order polynomial and we use this to subtract out the
foreground contribution from the entire range $ \Delta \nu \le \, 6
\times \Delta \nu_{0.1}$.  The value of $\Delta \nu_{0.1}$ and the
polynomial fit are both shown in Figure \ref{clnu}. Note that $\Delta
\nu_{0.1}$ decreases with increasing $\ell$, and it is less than $0.5
\, {\rm MHz}$ for all the $\ell$ values that we have considered.
Tests with simulations (Paper I) show that very little of the HI
signal is lost in this subtraction procedure.

 It is noteworthy that a variety of foreground removal techniques
 (e.g. \citealt{mcquinn,jelic,gleser,harker,Liu1,liu,nada}) all
 attempt to remove the foregrounds from images or visibilities {\it
   before} determining the power spectrum.  The HI signal is spread
 out along the entire frequency axis and all these techniques run the
 risk of removing a part of the HI signal along with the foregrounds.
 On the contrary, we have attempted foreground removal {\it after}
 determining the angular power spectrum. The signal, here, is
 localized in $\Delta \nu$ , and it is possible to reduce the risk of
 removing a part of the HI signal by suitably tuning the fitting
 procedure.

\begin{figure*}
\includegraphics[width=70mm,angle=-90]{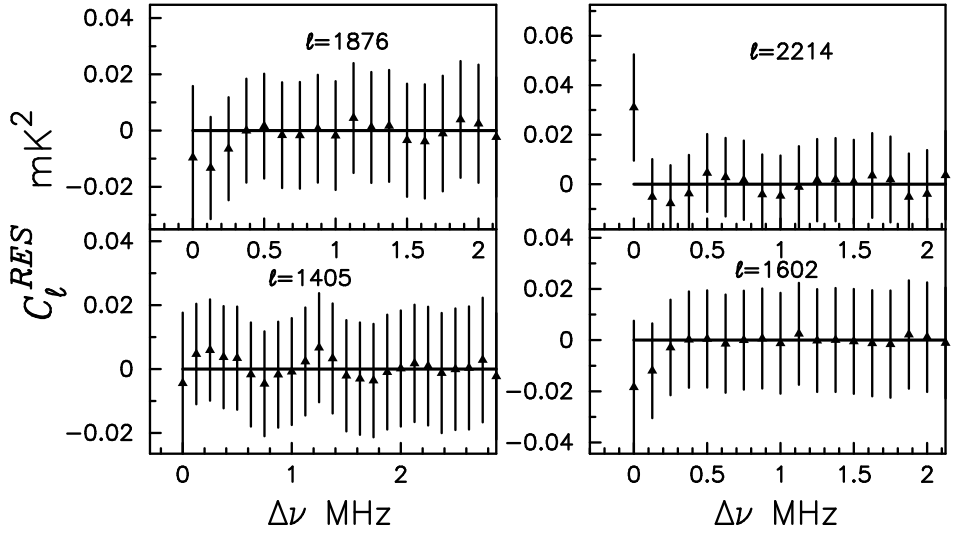}
\caption{The residuals  are shown  with $3\sigma$ error bars (system
  noise only).} 
\label{clres}
\end{figure*}

We expect the residual $C^{\rm RES}_{\ell}(\Delta \nu)$ that remains
after polynomial subtraction (Figure \ref{clres}) to contain only the
HI signal and noise provided the foregrounds have been successfully
removed.  In our observations the HI signal is much smaller than the
noise $\sigma \sim 0.01 \, {\rm mK}^2$.  We thus expect the residuals
to be consistent with noise provided the foregrounds have been
completely removed. We find that the residuals are consistent with
noise at the $3 \, \sigma $ level ({\it ie.}  $C^{\rm
  RES}_{\ell}(\Delta \nu) \le \, 0 \, \pm 3 \, \sigma$) at the three
smallest values of $\ell$ ($1405,1602,1876$). This establishes that
our foreground removal technique works.  This technique, however, is
not as successful at $\ell=2214$ where a single point of $\Delta \nu=0$
has a value that is somewhat larger than $0 \pm 3 \, \sigma$.  We have
also carried out the entire analysis for several other, larger, values
of $\ell$ for which the results have not been shown here.  While
sidelobe suppression works quite well in removing the oscillations, we
are unable to completely remove the foregrounds for the other $\ell$
values.

The fact that we have three $\ell$ values where the foregrounds have
been completely removed and $C^{\rm RES}_{\ell}(\Delta \nu)$ is
consistent with noise allows us to place an upper limit on the HI
signal.  We assume that the HI traces the dark matter with a possible
linear bias $b$ whereby $C^{\rm HI}_{\ell}(\Delta \nu)$ can be
calculated (Paper I) in terms of the dark matter power spectrum, and
the cosmological parameters for which we have used the values
$(\Omega_{m0},\Omega_{\Lambda0},h,\sigma_8,n_s)=(0.3,0.7,0.7,1.0,1.0)$.
The HI signal is now completely determined upto a proportionality
factor $C^{\rm HI}_{\ell}(\Delta \nu) \ \propto \ [\bar{x}_{\rm HI}
  b]^2$ where $\bar{x}_{\rm HI}$ is the mean neutral fraction of
hydrogen gas. We have considered $\bar{x}_{\rm HI} b$ as a free
parameter, and performed a likelihood analysis using our observational
data to place an upper limit on $\bar{x}_{\rm HI} b$. In our
analysis we have assumed that the $C_{\ell}(\Delta \nu)$ at the
different $\ell$ values are independent. For each $\ell$, we have only
used $C_{\ell}(\Delta\nu)$ in the range $\Delta \nu < \Delta
\nu_{0.1}$; the $\Delta \nu $ values that were used to estimate the
polynomial for foreground removal were excluded in the likelihood
analysis.  The covariance between $C_{\ell}(\Delta\nu)$ at different
$\Delta \nu$ values was taken into account in estimating the
likelihood. We place an upper limit $\bar{x}_{\rm HI} b \le 2.9$
with $99 \%$ confidence using our observational data. In Paper~I a
high-pass filter had been applied to remove the oscillatory pattern at
a single $\ell$ value (the smallest) to give an upper limit of $7.95$
for $\bar{x}_{\rm HI} b$. Table \ref{tab:parm} summarizes the
results of the current work, and presents a comparison with the
results of Paper~I.  It is clear that the present analysis is a
considerable improvement over the previous result. In conclusion we
show that sidelobe suppression can bring about a considerable
improvement in foreground removal.

\begin{table}
\caption{We show $\Delta C_{\ell}(\Delta \nu=0)$, the $1-\sigma$ error,
  for the $\ell$ values where it was  possible to successfully
  remove the foregrounds.  The upper limits on  $\bar{x}_{\rm HI} b$   
from the single $\ell$ value  in Paper I and from combining the three
different $\ell$ values in the current work are also shown. } 
\vspace{.2in}
\label{tab:parm}
\begin{tabular}{|p{2cm}|p{2cm}||p{3cm}|}
\hline
$\ell$&$\Delta C_{\ell}(\Delta \nu=0)$&$\bar{x}_{\rm HI} b$ \\
 &  ${\rm mK}^2$& $99\%$ confidence\\
\hline
(Paper I)& &\\
$1476$&$0.010$& $\le 7.95$\\
\hline
\hline
(Current work)& &\\
$1405$&$0.007$&\\
$1602$&$0.008$&$\le 2.9$\\
$1876$&$0.008$&\\
\hline
\end{tabular}
\end{table}

\section{Acknowledgment}

We would like to thank the anonymous referee for providing us with
constructive suggestions which helped to improve the paper. The
authors would like to thank Ravi Subrahmanyan for pointing out the
side-lobe effect during a workshop at HRI, Allahbad . AG thanks
S.P. Khastgir for a detailed reading of the manuscript. AG
acknowledges the Council of Scientific and Industrial Research (CSIR),
India for financial support through Senior Research Fellowship
(SRF). SSA would like to acknowledge CTS, IIT Kharagpur for the use of
its facilities and the Associateship Programme of IUCAA for providing
support. The data used in this paper was obtained using GMRT. The GMRT
is run by the National Centre for Radio Astrophysics of the Tata
Institute of Fundamental Research. We thank the GMRT staff for making
these observations possible.

\end{document}